\documentclass{aa}  
\usepackage{graphicx}
\usepackage{txfonts}
\def\degr{\hbox{$^\circ$}}
\def\arcmin{\hbox{$^\prime$}}
\def\arcsec{\hbox{$^{\prime\prime}$}}
\begin{document}
   \title{Probing the formation of intermediate- to high-mass stars in protoclusters}

   \subtitle{II. Comparison between millimeter interferometric observations of NGC~2264-C and SPH simulations of a collapsing clump}

\titlerunning{High-mass star formation in NGC~2264-C}

   \author{N. Peretto
          \inst{1, 4}
          P. Hennebelle\inst{2}
	\and
	P. Andr\'e\inst{1}\inst{,3}
          }


   \institute{Service d'Astrophysique, CEA/DSM/DAPNIA, C.E. Saclay, Orme des Merisiers, 91191 Gif-sur-Yvette, France
         \and
            Laboratoire de radioastronomie millim\'etrique, UMR 8112 du CNRS, \'Ecole normale sup\'erieure et Observatoire de Paris, 24 rue Lhomond, 75231 Paris Cedex 05, France
\and
 AIM -- Unit\'e Mixte de Recherche CEA -- CNRS -- Universit\'e Paris VII -- UMR 7158, France
 \and
 Department of Physics  \& Astronomy, University of Manchester, P.O. Box 88, Manchester M60 1QD, United Kingdom}   

 \date{Submitted 22 May 2006 / Accepted 30 October 2006}


  \abstract
{}
 {The earliest phases of massive star formation in clusters are still poorly understood. Here, we test the hypothesis for high-mass star formation proposed in our earlier paper (Peretto et al. 2006) stating that a massive, ultra-dense core may be currently 
 forming at the center of the collapsing NGC~2264-C protocluster via the gravitational coalescence of several intermediate-mass Class 0 objects. }
  {In order to confirm the physical validity of this hypothesis, we carried out IRAM Plateau de Bure interferometer observations of NGC~2264-C 
  and performed SPH numerical simulations of the collapse of a Jeans-unstable, prolate dense clump. A detailed comparison between these 
  hydrodynamic simulations and both our earlier IRAM 30m observations and the new interferometer observations is presented.} 
 {Our Plateau de Bure observations provide evidence for disk emission in three of the six Class~0-like objects identified earlier with the 
 30m in the NGC~2264-C clump. Furthermore, they reveal the presence of a new compact source (C-MM13) located only $\sim 10000$~AU away, 
 but separated by $\sim 1.1$~km.s$^{-1}$ in (projected) velocity, from the most massive Class~0 object (C-MM3) lying at the very center of NGC~2264-C.
 Detailed comparison with our numerical SPH simulations supports the view that NGC~2264-C is an elongated cluster-forming clump in the 
 process of collapsing and fragmenting along its long axis, 
 leading to a strong dynamical interaction and possible protostar merger in the central region of  the clump. 
 The marked velocity difference observed between the two central objects C-MM3 and C-MM13, which can be reproduced in the simulations, 
 is interpreted as an observational signature of this dynamical interaction.
 The present study also sets several quantitative constraints on the initial conditions of large-scale collapse in NGC~2264-C. 
 Our hydrodynamic simulations indicate that the observed velocity pattern characterizes an early phase of protocluster collapse which survives 
 for an only short period of time  (i.e., $\le1\times 10^5$~yr). To provide a good match to the observations the simulations require an initial ratio of turbulent to 
 gravitational energy of only $\sim 5\, \%$, which strongly suggests that the NGC~2264-C clump is structured primarily by gravity rather than turbulence. 
 The required ``cold'' initial conditions may result from rapid compression by an external trigger.  }
  {We speculate that NGC~2264-C is not an isolated case but may point to key features of the initial phases of high-mass star formation in protocluster.}

   \maketitle
%

\section{Introduction}

Most stars are believed to form in clusters (e.g. Lada \& Lada 2003) and high-mass stars 
may form exclusively in cluster-forming clouds. 
For a comprehensive understanding of clustered star formation, a good knowledge of
the initial conditions and earliest phases of the process is crucial which, 
in practice, can only be inferred from detailed studies of deeply embedded protoclusters 
at (sub)millimeter wavelengths (e.g. Motte et al. 1998, Andr\'e 2002).

Two main scenarios have been proposed to explain the formation of 
high-mass stars in clusters.
In the first scenario, high-mass stars form essentially in the same way as low-mass stars, via an enhanced accretion-ejection phase. 
In the standard model of low-mass star formation, the mass accretion rate is governed by the thermal sound speed and does not exceed $\sim 10^{-5}$~M$_{\odot}$.yr$^{-1}$ (e.g. Shu 1977; Stahler et al. 2000) in cold cores. To form high-mass stars by accretion,  
a significantly higher accretion rate $\sim 10^{-3}$~M$_{\odot}$.yr$^{-1}$ is required to overcome the radiation pressure generated 
by the luminous central object (e.g. Wolfire \& Cassinelli 1987).
In order to solve this problem, McKee \& Tan (2003) proposed a model in which high-mass star formation takes place in 
ultra-dense cores supported by turbulence within virialized cluster-forming clumps. This model produces high-mass accretion rates 
such as those required to form high-mass stars by accretion.

In the second scenario, high-mass stars form by coalescence of lower-mass stars in the dense inner core 
of a contracting protocluster (Bonnell et al. 1998). 
This scenario requires high stellar densities (i.e. $\sim 10^{8}$~stars.pc$^{-3}$) 
in order to render the probability of stellar collisions high enough and allow stellar mergers to take place.
It avoids the accretion problem of high-mass star formation by directly combining  
the masses of lower, intermediate-mass stars. However, no detailed model 
exists yet to describe how this coalescence mechanism actually occurs.

\begin{table*}
\begin{minipage}[!ht!]{\textwidth}
\caption{Measured source properties}        
\label{resume_C1}      
\renewcommand{\footnoterule}{}  
\begin{tabular}{l c c c c c c c c}        
\hline\hline                 
Source     
& Coordinates & Undec.FWHM & P.A.  & S$_{peak}^{1.2}$&S$_{peak}^{30m}$ & S$_{peak}^{exp:3.2}$ & S$_{peak}^{3.2}$& $S_{int}^{3.2}$ \\
& ($\alpha _{2000}$ $\,$ $\delta _{2000}$) & (arcsecond) & (deg)  & (mJy/beam)& (mJy/beam) & (mJy/beam) & (mJy/beam) & (mJy) \\
 \hspace{0.3cm}[1] & [2] & [3] & [4] & [5] & [6] & [7] & [8] & [9]  \\    
\hline
C-MM1	  & 06:41:17.95 $\,$ +09:29:03  &  5.8$\times$4.1& 63 & -- &255 & 6 & 18 & 21
\\
C-MM2     & 06:41:15.15 $\,$ +09:29:10 & 5.6$\times$4.3 & 50  &-- &183 & 4 & 7 &  8
\\
C-MM3    & 06:41:12.30 $\,$ +09:29:12 & 5.7$\times$4.4 & 58 &224& 573 & 14  & 37 & 45
\\
C-MM4     & 06:41:09.95 $\,$ +09:29:22 & 7.0$\times$5.0 & 87  &77& 426& 10 & 14  & 24
\\
C-MM5    &  06:41:10.15 $\,$ +09:29:36 & 6.2$\times$4.5 & 83  & --&261 & 6 &  5 &  7
\\
C-MM9    &  06:41:15.30 $\,$ +09:29:33 & 7.3$\times$4.3  & 51  &--& 94 & 2 &  6 & 9
\\
C-MM13   & 06:41:11.45 $\,$ +09:29:17 & 7.5$\times$6.1 & 87   & --&--&--&  5 & 11
\\
\hline 

\end{tabular}
\flushleft
[1]: The C-MM numbers are the same as in PAB06. The new source is labelled C-MM13.

[2]: J2000 source coordinates, accurate to better than 1\arcsec, derived from a Gaussian fit 
to the PdBI 3.2mm dust continuum map.

[3]: Undeconvolved FWHM sizes derived from fitting an elliptical Gaussian to the PdBI 3.2mm dust continuum map.

[4]: Position angle (from North to East) of the major axis of the fitted Gaussian ellipse from the 3.2~mm dust continuum map.

[5]: PdBI 1.2mm peak flux density at the best-fit source position (HPBW=3.1\arcsec$\times$1.5\arcsec)

[6]: 30m 1.2mm peak flux density at the source position (HPBW=11\arcsec ; from PAB06)

[7]: 3.2mm peak flux density expected at PdBI angular resolution estimated from col.[6] (HPBW=4.5\arcsec)

[8]: PdBI 3.2mm peak flux density at the best-fit source position (HPBW=5.3\arcsec$\times$3.8\arcsec)

[9]: PdBI 3.2mm integrated flux density inside the fitted Gaussian ellipse

\end{minipage}                      
\end{table*}

On the observational side, studying the earliest stages of high-mass star formation is 
particularly difficult due to the tight packing, deeply embedded nature, 
and relatively large distances of massive protoclusters. 
Based on IRAM 30m observations of the massive cluster-forming clump NGC~2264-C (d$\sim 800$~pc), 
Peretto, Andr\'e, Belloche (2006 -- hereafter PAB06) recently proposed a picture of high-mass star formation combining 
features of the two above-mentioned scenarios. 
They showed that NGC~2264-C harbored a dozen Class~0-like objects (cf. Andr\'e, Ward-Thompson, Barsony 2000) and 
was characterized by large-scale collapse motions (see also Williams \& Garland 2002).
They suggested that a massive, ultra-dense protostellar core was in the making in the central part of the 
NGC~2264-C clump as a result of the gravitational merger of two or more lower-mass Class~0 objects. 
The total mass inflow rate associated with the collapse of the clump toward 
the central protostellar core was estimated to be $3 \times 10^{-3}$~M$_{\odot}$.yr$^{-1}$. 
PAB06 argued that the combination of large-scale collapse and protostellar mergers may be the key 
to produce the conditions required for high-mass star formation in the center of NGC~2264-C. 

In this paper, we follow up on the detailed single-dish study of NGC~2264-C
by PAB06 and present higher-resolution observations of the same cluster-forming
clump taken with the IRAM Plateau de Bure interferometer. 
We compare our observations with a set of SPH hydrodynamic numerical simulations 
which attempt to specifically model NGC~2264-C. 
As the kinematical and density patterns of NGC~2264-C appear to be relatively
simple, comparison between millimeter observations of this region and numerical models 
offers a unique opportunity to make progress in our understanding of clustered star formation.

Section~2 presents our PdBI observations. 
Section~3 describes the dedicated hydrodynamic SPH simulations that we performed to model NGC~2264-C. 
We compare the observations with the numerical simulations in Sect.~4  and draw several conclusions in Sect.~5.

\section{Interferometer observations of NGC~2264-C}

\subsection{Observations}

We performed 3.2~mm and 1.2~mm observations of the central part of NGC~2264-C with the IRAM 
Plateau de Bure interferometer (PdBI) in December 2003 and April 2004. We used the C and D configurations with 6 antennas. We used both 1~mm and 3~mm receivers with 244.935620~GHz ($\lambda = 1.2$~mm) and 93.176258~GHz ($\lambda = 3.2$~mm) as central rest frequencies. We observed at four positions which were chosen so as to obtain a fully sampled mosaic at 3.2~mm (primary beam FWHM $\sim 54$\arcsec) 
and to encompass the millimeter sources C-MM1, C-MM2, C-MM3, C-MM4, CMM5, and C-MM9 (see Fig.~2b of PAB06) identified by PAB06 with the IRAM 30m telescope. Because the corresponding 1.2~mm mosaic is undersampled (primary beam FWHM $\sim 20$\arcsec), only two of 
these sources (C-MM3 and C-MM4) were effectively imaged at 1.2mm. We obtained a 3.2~mm dust continuum mosaic and two separate 
1.2~mm continuum maps,  as well as a N$_2$H$^+$(1-0) mosaic.
The spectral resolution for the N$_2$H$^+$(1-0) data was 20~kHz, which corresponds to a velocity resolution of 0.06~km.s$^{-1}$ at 93.2~GHz.
The sources used for the bandpass, amplitude and phase calibrations were 
0420-014, 0528+134, 0736+017, CRL618, 0923+392, and 3C273 
(only the first three were used for the second run in April). We calibrated the data and 
produced images using the CLIC and MAPPING softwares (Lucas 1999, Guilloteau et al. 2002), part of the GILDAS package\footnote{See \texttt{http://www.iram.fr/IRAMFR/GILDAS} for more information about the \textsc{gildas} softwares.} (Pety 2005). 
The deconvolution was performed using the natural weighting option of the Clark (1980) 
CLEAN algorithm (Guilloteau 2001).
The final synthesized beam was 5.3\arcsec$\times$3.8\arcsec ~ (HPBW) with P.A.=+63\degr ~ 
at 3.2~mm, and 3.1\arcsec$\times$1.5\arcsec ~(HPBW) with P.A.=+74\degr ~ at 1.2~mm. 
We also combined our PdBI N$_2$H$^+$(1-0) observations with the single-dish  N$_2$H$^+$ data cube of PAB06 in order to recover 
short-spacing information. 
The resulting synthesized beam of the combined N$_2$H$^+$(1-0) mosaic is 
6.1\arcsec$\times$4.0\arcsec ~(HPBW) with P.A.=+65\degr.

\subsection{Dust continuum results: Evidence for disk emission}

\begin{figure}[t]
\hspace{-0.5cm}
\includegraphics[height=10cm,angle=270]{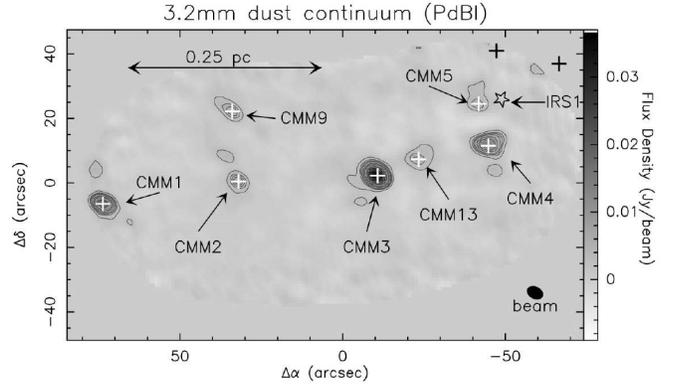}
\caption{3.2~mm dust continuum mosaic of the central part of NGC~2264-C obtained with the PdBI. The (0\arcsec, 0\arcsec) position corresponds to ($\alpha = 06^h13^m00^s$; $\delta = 09^o29\arcmin10\arcsec$) in J2000 coordinates. The pre/protostellar sources detected in this map are marked with white crosses and labelled C-MM, keeping the same numbering scheme as in PAB06. The two black crosses mark the positions of C-MM10 and C-MM12 (cf. PAB06), which are located at the very edge of our mosaic. 
The black open star shows the position of the luminous $IRAS$ source IRS1. 
The rms of the mosaic is $\sigma \simeq 0.8$~mJy/beam. 
The contours are 2.4~mJy/beam (i.e. 3$\sigma$), 4~mJy/beam (i.e. 5$\sigma$), 5, 6, 8~mJy/beam and then go from 10 to 35~mJy/beam by 5~mJy/beam. 
\label{n2264c_3mm}}
\end{figure}

Our PdBI 3.2~mm dust continuum mosaic is shown in Fig.~\ref{n2264c_3mm}. It 
reveals only pointlike sources, since most of the extended emission seen in the single-dish 1.2~mm dust 
continuum map (cf Fig.~2 of PAB06) was filtered out by the interferometer. 
The final rms noise level was $\sigma \sim~0.8$~mJy/beam. We extracted millimeter sources from this map using the Gaussclump algorithm (Stutzki \& G\"usten 1990). We detected seven peaks lying above 5$\sigma$. All of these peaks were previously detected with the 30m telescope, except one object,  here called C-MM13 (cf. Fig.~1),  which is a new detection. Several other peaks lie between 3$\sigma$ and 5$\sigma$, but, by lack of confidence in these marginal detections, we did not consider them. 
The present 3.2~mm continuum map confirms and improves the positions of the compact millimeter continuum sources detected by PAB06. Interestingly our PdBI 1.2~mm continuum observations of C-MM3 and C-MM4 (not shown here) do not reveal any further sub-fragmentation. The source properties as derived from our PdBI and 30m observations are summarized in Table~\ref{resume_C1}.

The millimeter dust continuum emission from a (Class~0 or Class~I) protostar a priori originates from two components: an extended envelope (a few thousand AU in size) and a disk (up to a few hundred AU). At a distance of 800~pc, the linear resolution of the 30m telescope is $\sim 9000$~AU (HPBW) at $\lambda = 1.2$~mm.
On this spatial scale, the 1.2mm continuum emission observed toward a young protostar 
(Table \ref{resume_C1} col.[6]) is expected to be dominated by the envelope rather than by the disk (e.g. Andr\'e, Ward-Thompson, Barsony 2000;
Looney, Mundy, Welch 2000).   Conversely, we expect the disk component to dominate on compact, interferometric scales
(e.g. Terebey et al. 1993). Assuming an isothermal, centrally-condensed envelope, i.e. with a density $\rho \propto r^{-2}$, the flux density is expected to scale linearly with beam size: S$_{\nu} \propto \theta$. 
We thus expect the envelope contribution to the PdBI flux density at 1.2~mm to be given by: 
\begin{equation}
S_{peak}^{exp:1.2} = S_{peak}^{30m} \times \left(\frac{HPBW_{Bure}}{HPBW_{30m}}\right)
\label{flux_1}
\end{equation}

If we assume the Rayleigh-Jeans regime and adopt a dust opacity scaling as $\kappa_{\nu} \propto \nu^{\beta}$ (e.g. Hildebrand 1983), then we can also estimate the expected contribution of the envelope to the PdBI flux density at 3.2~mm:

\begin{equation}
S_{peak}^{exp:3.2} = S_{peak}^{exp:1.2} \times \left(\frac{1.2}{3.2}\right)^{\beta+2}
\label{flux_2}
\end{equation}

 In order to estimate a lower limit to the disk component, we choose $\beta=1$ which maximizes the contribution of the envelope.
 A value of $\beta=1.5$ is likely to be more representative of protostellar cores/envelopes (e.g. Ossenkopf \& Henning 1994) and 
 would yield a lower estimate for the expected envelope contribution. The expected envelope flux densities are listed in Table \ref{resume_C1} col.[7] for each detected source. It can be seen that they are a factor of $\sim 2-3$
 lower than the observed flux densities (Table \ref{resume_C1} col.[8]) for C-MM1, C-MM3 and C-MM9. The excess flux density observed on small spatial scales can be attributed to unresolved disk emission (e.g. Terebey et a. 1993). 
 Our results thus suggest the presence of a disk in C-MM1, C-MM3, C-MM9 and confirm the protostellar nature of these 
 candidate Class~0 sources. Given the uncertainties on the dust emissivity index $\beta$ (e.g. Dent et al. 1998), we cannot conlude on the presence or absence of a disk in C-MM2 and C-MM4. Finally, it is very likely that C-MM5 does not have a disk since its observed 3.2~mm  flux density 
is consistent with pure envelope emission.

For the three sources showing evidence of disk emission, i.e. C-MM1, C-MM3 and C-MM9, we can estimate both the disk and envelope masses as follows. 
First, we estimate the flux arising from the disk by subtracting the expected envelope peak flux density given in Table~\ref{resume_C1} col.[7] 
from the observed peak flux density of col.[8] (here, we assume that our 3.2~mm PdBI observations do not spatially resolve the disk given the distance of NGC~2264). 
Then, the flux arising from the envelope is considered to be given by the integrated flux of col.[9] minus the disk contribution. 
These flux estimates are listed in Table~\ref{flux_diskenv}. For the envelope mass estimates, we follow PAB06 and assume a dust temperature T$_d = 15$~K, $\beta=1.5$, as well as a dust opacity $\kappa_{1.2mm}= 0.005$~cm$^{2}$.g$^{-1}$ corresponding to $\kappa_{3.2mm}=1.3\times 10^{-3}$~cm$^{2}$.g$^{-1}$.
Concerning the disk mass estimates, we assume a dust temperature of T$_d = 50$~K and a dust opacity $\kappa_{1.2mm}= 0.02$~cm$^{2}$.g$^{-1}$ (Beckwith et al. 1990) corresponding to $\kappa_{3.2mm}=5.2\times 10^{-3}$~cm$^{2}$.g$^{-1}$ (see Table~\ref{flux_diskenv}). 
The dust temperature is higher for the disk because it is supposed to be warmer, closer to the star, while the dust opacity is 
slightly different from the one adopted in PAB06 because of the enhanced dust emissivity expected in the dense central parts of protostellar disks 
(e.g. Ossenkopf \& Henning 1994). We caution that the disk masses calculated in this way are only rough estimates.
A more proper analysis of the density structure of these sources, especially C-MM3, through sub-arcsecond millimeter observations would be of great interest. 
By deconvolving the FWHM sizes of Table \ref{resume_C1} from the synthesized beam, we derive the geometrical mean diameter of each source. 
For each object, we can then estimate the mean column density and mean volume density of the envelope component.
All of these derived source parameters are listed in Table~\ref{resume_mass}. 
It can be seen that C-MM3 and C-MM4 have the most massive envelopes by far, with $M_{env } \ge 15\, M_{\odot}$ in both cases.
The densest envelope/core is associated with the central source C-MM3, which reaches a mean volume density $\ge$ 1$\times$10$^8$~cm$^{-3}$ 
on a $\sim 3200$~AU (2$\times$FWHM) scale.

\begin{figure}[t]
\hspace{0.0cm}
\includegraphics[height=8cm,angle=270]{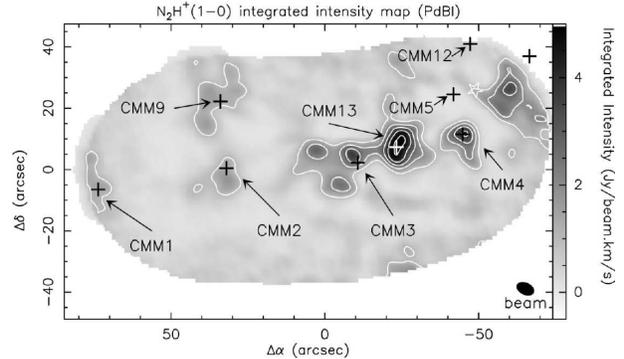}
\vspace{-3cm}
\caption{N$_2$H$^+$(1-0) integrated intensity map of the central part of NGC~2264-C obtained with the PdBI. The (0\arcsec, 0\arcsec) position corresponds 
to ($\alpha = 06^h13^m00^s$; $\delta =09^o29\arcmin10\arcsec$) in J2000 coordinates. 
The crosses with labels show the positions of the pre/protostellar cores detected in our PdBI dust continuum mosaic (Fig.~\ref{n2264c_3mm}), 
while the black cross without label mark the position of C-MM10 (cf. PAB06). The white open star shows the position of the 
$IRAS$ source IRS1. The contours go from 1 to 5 Jy/beam.km/s by 1 Jy/beam.km/s.
\label{n2264c_n2h+_pdbi}}
\end{figure}

The new source identified with the interferometer, C-MM13, could not be separated from C-MM3 at the angular resolution of the IRAM 30m telescope. The projected distance between C-MM3 and C-MM13 is only $\sim 10000$~AU. 
Surprisingly, C-MM13 is the only new compact millimeter continuum source detected with the PdBI above 5$\sigma$.  
The fact that there is almost no sub-fragmentation despite a factor 2-4 
improvement in angular resolution between the partially resolved 30m sources and their PdBI counterparts suggests that most of the compact dust continuum sources detected at the 30m represent individual Class~0 objects
rather than small groups of protostellar cores. However, we stress that our PdBI observations could only detect 
dust continuum sources with 3.2~mm peak flux densities larger than 4~mJy/beam (i.e. 5$\sigma$). 
Assuming the same temperature and dust properties as above for the envelope mass estimates, this corresponds to a mass detection threshold $\sim 3$~M$_{\odot}$. 
Therefore, we could not have detected low-mass pre/protostellar cores possibly lying in the vicinity of the main sources listed in Table~\ref{resume_C1}. 
This point is further discussed in the next section.

\begin{table}
\begin{minipage}[ht]{\columnwidth}
\caption{Estimated PdBI 3.2~mm flux densities of the disk and envelope components for the three objects showing evidence of disk emission}        
\label{flux_diskenv}      
\centering                          
\renewcommand{\footnoterule}{}  
\begin{tabular}{c c c c c c }      
\hline\hline                 
Source   & S$_{disk}$\footnote{3.2~mm flux density of the disk component estimated by subtracting the peak flux density expected for the envelope 
(col.[7] of Table~\ref{resume_C1}) from the observed peak flux density (col.[8] of Table~\ref{resume_C1}) }& S$_{peak}^{env}~$\footnote{Estimated 
3.2~mm peak flux density of the envelope component at the PdBI resolution (cf. col.[7] of Table~\ref{resume_C1}) }& S$_{int}^{env}$\footnote{3.2~mm integrated 
flux density of the envelope component estimated by subtracting the disk contribution given in col.[2] from the total integrated flux density measured with 
PdBI (cf.  col.[9] of Table~\ref{resume_C1}) } & M$_{disk}$\footnote{Disk mass seen with the PdBI and estimated from S$_{disk}$ (with T$_d = 50$~K and $\kappa=5.2\times10^{-3}$~cm$^2$.g$^{-1}$). Typical uncertainty is a factor $\ge 2$ (on either side) due to uncertain dust opacity and dust temperature.}     \\
& (mJy) & (mJy/beam) &(mJy) & (M$_{\odot}$) \\    
\hline
C-MM1	&  12 & 6 & 9 & 0.6
\\ 
C-MM3    &  23 & 14 & 22 & 1.1
\\
C-MM9    & 4 & 2 & 5 & 0.2
\\
\hline
\end{tabular}
\end{minipage}                           
\end{table}

\begin{table*}
\begin{minipage}[ht]{\textwidth}
\caption{Derived source parameters}        
\label{resume_mass}      
\centering                          
\renewcommand{\footnoterule}{}  
\begin{tabular}{ c c c c c c c}       
\hline\hline                 
Source   & FWHM\footnote{Geometrical mean of the deconvolved, major and minor FWHM diameters measured on the PdBI  3.2mm continuum mosaic } 
& N$_{H_2}$\footnote{Column density of the envelope seen with the PdBI and estimated from  S$_{peak}^{3.2}$ for the sources without a disk (col.[8] of Table~\ref{resume_C1}) and from S$_{peak}^{env}$ for the sources with a disk (col.[3] of Table~\ref{flux_diskenv}). The used dust properties are T$_d = 15$~K and $\kappa=1.3\times10^{-3}$~cm$^2$.g$^{-1}$. Typical uncertainty is a factor $\ge 2$ (on either side) due to uncertain dust opacity and dust temperature.}  & M$_{env}$\footnote{Envelope mass seen with the PdBI and estimated from S$_{int}^{env}$ (Table~\ref{flux_diskenv}; with T$_d = 15$~K and $\kappa=1.3\times10^{-3}$~cm$^2$.g$^{-1}$). The uncertainty is the same as for the column density.}   & M$_{core}^{30m}$\footnote{Core mass seen with the 30m telescope (from PAB06). This mass is larger than M$_{env}$ because, contrary to the PdBI, the 30m does not filter out most of the large scale emission} &n$_{H_2}$\footnote{Envelope volume density estimated in a radius equal to the FWHM. Estimated in a radius twice smaller, the density increases by a factor of 4. The uncertainty is the same as for the column density}  & V$_{LSR}$\footnote{LSR velocity estimated from the combined N$_2$H$^{+}$(1-0) spectra}       \\
&  (AU) & (10$^{23}$~cm$^{-2}$)  &  (M$_{\odot}$) & (M$_{\odot}$) & (cm$^{-3}$) &(km.s$^{-1}$) \\    
\hline
C-MM1	&  1400 &  7 & 6.2 & 13.1&8.3$\times$10$^7$&--
\\ 
C-MM2     & 1400 &  9  & 5.5 & 16.0 &7.3$\times$10$^7$& 6.2
\\
C-MM3    &1600 &  17  & 15.2 &  40.9&1.4$\times$10$^8$& 7.1
\\
C-MM4     & 3000& 17   & 16.5 &  35.1&2.2$\times$10$^7$ & 8.9
\\
C-MM5    &  2200 & 6   & 4.8 & 18.4&1.6$\times$10$^7$ &--
\\
C-MM9    & 2400  &  2   & 3.5  &  6.6&9.3$\times$10$^6$&--
\\
C-MM13   &  4000 &  6  & 7.6  & -- &4.3$\times$10$^6$ & 8.2
\\
\hline
\end{tabular}
\end{minipage}                           
\end{table*}

\subsection{N$_2$H$^+$(1-0) results}
\subsubsection{PdBI observations only}
Our  PdBI N$_2$H$^+$(1-0) integrated intensity map is shown in Fig.~\ref{n2264c_n2h+_pdbi}. As for the dust continuum mosaic, most of the extended emission is filtered out.
On this map, it can be seen that C-MM5 and possibly C-MM12 are not closely associated with a N$_2$H$^+$(1-0) peak. We also note that the strongest N$_2$H$^+$(1-0) peak is associated with the new C-MM13 object which is one of the weakest dust continuum sources. The morphological differences between the dust continuum sources and their N$_2$H$^+$(1-0) counterparts may possibly reflect differences in chemical evolutionary stage (cf. Aikawa et al. 2005). It is also noteworthy that several N$_2$H$^+$(1-0) peaks  
are not associated with any of the bona fide dust continuum sources listed in Table~\ref{resume_mass}. 
On the other hand, most of these N$_2$H$^+$(1-0) peaks have faint dust continuum counterparts with flux levels between 3$\sigma$ and 5$\sigma$ in the 
3.2~mm continuum map (compare Fig.~\ref{n2264c_3mm} and Fig. ~\ref{n2264c_n2h+_pdbi}).

The  N$_2$H$^+$(1-0) spectra observed at PdBI (prior to combination with 30m data) provide an estimate of the velocity dispersion within the sources on a 
$\sim 3500$~AU (FWHM) spatial scale. The mean line-of-sight velocity dispersion is found to be $<\sigma_{los}>\simeq 0.34$~km.s$^{-1}$. 
Assuming a kinetic temperature T$_k = 15$~K, the non-thermal contribution to this velocity dispersion is $<\sigma_{los}^{NT}> \simeq 0.33$~km.s$^{-1}$. 
Comparing this value with the isothermal sound speed, c$_s \simeq 0.23$~km.s$^{-1}$, we conclude that the sources are still marginally dominated by non-thermal motions 
(due to, e.g,  turbulence, collapse or, outflow) on scales of a few thousand AUs.

\subsubsection{Combined PdBI and 30m observations}

\begin{figure}[!t!]
\hspace{0.0cm}
\includegraphics[height=8cm,angle=270]{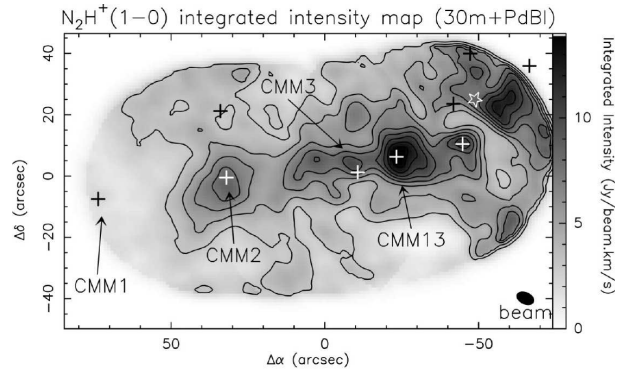}
\vspace{-3cm}
\caption{N$_2$H$^+$(1-0) integrated intensity map of the central part of NGC~2264-C resulting from the combination of our 30m and PdBI data. The (0\arcsec, 0\arcsec) position corresponds to ($\alpha = 06^h13^m00^s$; $\delta = 09^o29\arcmin10\arcsec$) in J2000 coordinates. The crosses mark the positions of the pre/protostellar 3.2~mm dust continuum sources detected in Fig.~\ref{n2264c_3mm}. For the sake of readability, only a few sources are labeled. The white open star symbol shows the position of the 
$IRAS$ source IRS1. The contours go from 3 to 13 Jy/beam.km/s by 2 Jy/beam.km/s. 
\label{n2264c_n2h+}}
\end{figure}

\begin{figure}[!t!]
\hspace{0cm}
\includegraphics[height=7cm,angle=270]{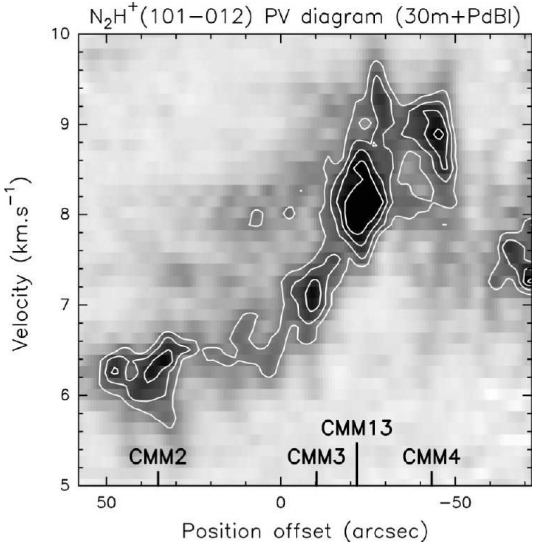}
\caption{Position-velocity diagram derived from the combined (PdBI~$+$~30m) N$_2$H$^+$(101-012) data cube 
by taking a cut along the East-West axis going through C-MM2, C-MM3, C-MM4 and C-MM13. 
The positions of these sources are plotted on the position axis. The white contours go from 0.5 to 1.1 Jy/beam by 0.2 Jy/beam. 
\label{pv_n2h+}}
\end{figure}

As mentioned in \S~2.1, we added short-spacing 30m information to our interferometric data in order to obtain a N$_2$H$^+$ map 
sensitive to a wide range of angular scales from $\sim 4\arcsec$ up to $\sim 4\arcmin$. 
This combination was performed using the MAPPING package developed by IRAM (Guilloteau et al. 2002). 
The combined PdBI/30m map of N$_2$H$^+$(1-0) integrated intensity is shown in Fig.\ref{n2264c_n2h+}. 
As expected, more extended emission is present in the combined mosaic, but the compact sources detected in the PdBI map are still clearly visible. 

In order to constrain the kinematical pattern of these sources within the NGC~2264-C clump, we constructed a position-velocity (PV) diagram 
along an East-West axis going through the four central sources, C-MM2, C-MM3, CMM4, and C-MM13 (see Fig.~\ref{pv_n2h+}). This PV diagram shows 
an overall velocity gradient of 8.4~km.s$^{-1}$.pc$^{-1}$ from East to West between C-MM2 and C-MM4. The LSR velocities of each of 
the four sources C-MM2, C-MM3, C-MM4, and C-MM13 are listed in Table~\ref{resume_mass}. Figure \ref{pv_n2h+} helps to clarify  
the origin of the velocity discontinuity identified by PAB06 with the 30m telescope in the center of NGC~2264-C (see Fig.~6 of PAB06). 
At the 30m resolution, the N$_2$H$^+$(101-012) spectrum observed toward the central source C-MM3 was double-peaked. 
The higher resolution of the PdBI interferometer now allows us to identify a distinct component, C-MM13, separated by 13\arcsec 
~in position and $\sim 1.1$~km.s$^{-1}$ in velocity from C-MM3. With the 30m telescope, C-MM13 could not be separated from C-MM3. We also observe in the western part of the PV diagram (at an offset of $-70$\arcsec~ and velocity of 7.4~km.s$^{-1}$) a velocity feature which is associated with the strong N$_2$H$^+$(1-0)  peak lying in this part of the clump (cf Fig.~\ref{n2264c_n2h+}). This velocity feature clearly departs from the rest of the diagram. We attribute it to a peculiar velocity field around 
the luminous young star IRS1, whose wind has likely perturbed the ambient velocity field and triggered star formation in the immediate vicinity 
($\sim 10$\arcsec~ in radius around IRS1), as suggested by the observations of Nakano et al. (2003) and Schreyer et al. (2003).

To summarize, our interferometric observations confirm the (Class~0) protostellar nature of C-MM1, C-MM3, C-MM9,  
and set new constraints on the kinematics of NGC~2264-C. 
The PdBI observations can help us to confirm or refute the scenario proposed by PAB06 of an axial collapse of NGC~2264-C along its long axis 
leading to the merging of dense cores in the center. 
In the context of this scenario, the protostellar nature of the  millimeter continuum 
cores sets strong timescale constraints: the individual collapse of the cores must occur on a significantly shorter timescale 
than the larger-scale collapse of the clump as a whole. 
The presence of two central sources, C-MM3 and C-MM13, adjacent to one another (i.e. 10000~AU) 
and with a velocity difference of $\sim 1.1$~km.s$^{-1}$, must also be accounted for. 
In the next two sections, we attempt to match these observational constraints with hydrodynamic simulations.

\section{SPH numerical simulations}

\subsection{Numerical method and initial conditions}

\begin{figure*}[th!]
\hspace{0cm}
\includegraphics[height=18cm,angle=0]{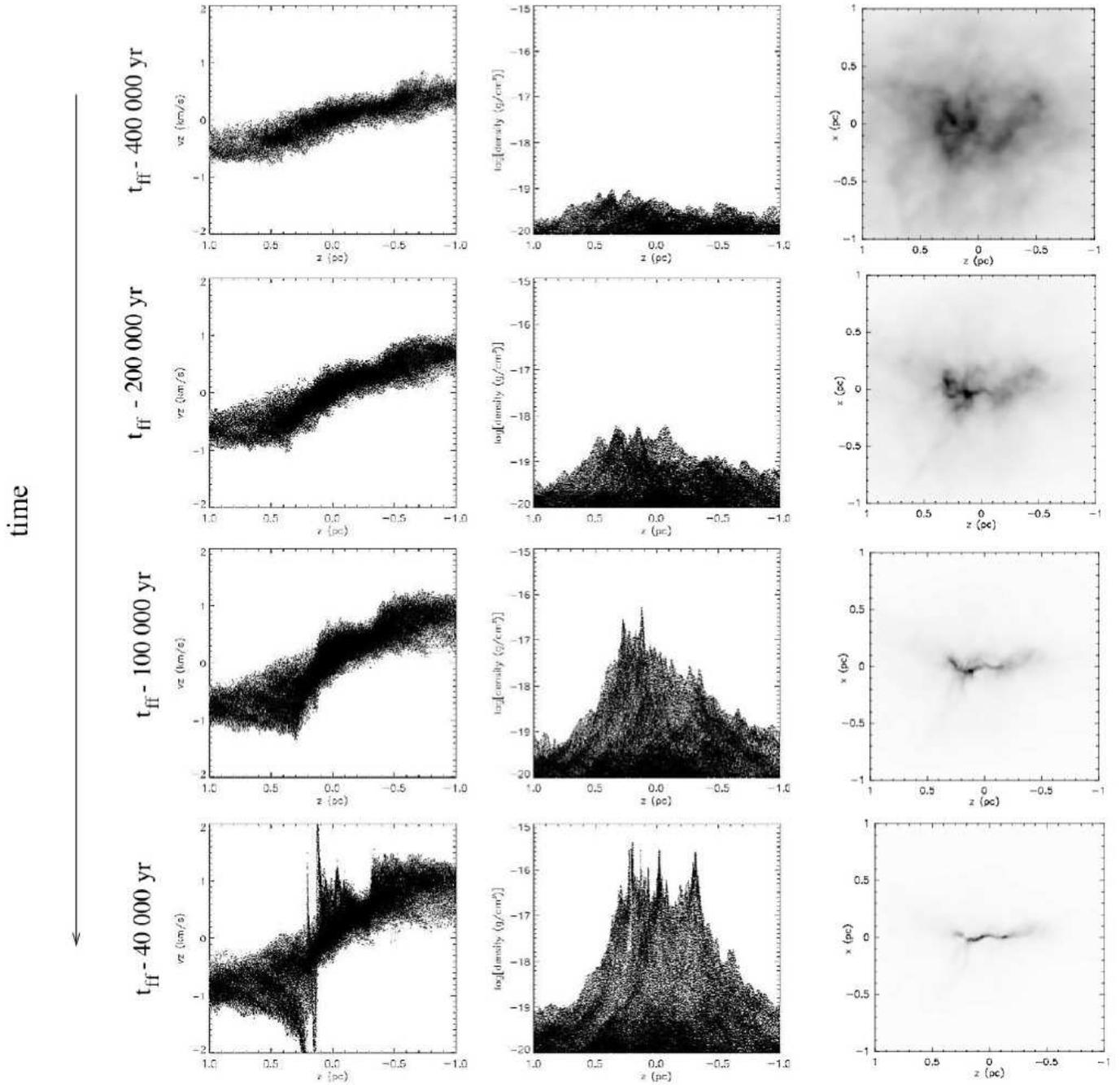}
\vspace{-0cm}
\caption{Time evolution of a simulation with an initial turbulent to gravitational energy ratio $\alpha_{turb}^0=5\%$. 
The first column displays the velocity field of the particles taken along the long axis (z-axis) of the model filament, as 
traced by individual SPH particles. The second column similarly displays the evolution of the density cut along the z-axis of the model filament.
The third column shows synthetic column density maps in the (z,x) plane. 
The reference time, labelled t$_{ff}$, is taken to be one global free-fall time of the initial clump after the start of the simulation, i.e., t$_{ff}=9.5\times10^5$~yr. 
The first row is taken at a time step  t$_{ff}-4\times10^5$~yr, while the second, third, and fourth rows are for time steps  t$_{ff}-2\times10^5$~yr, 
t$_{ff}-1.0\times10^5$~yr, and t$_{ff}-0.4\times10^5$~yr, respectively.
\label{sim}}
\end{figure*}

In order to test the physical validity of the scenario proposed by PAB06 (see also \S ~1), we performed
Smooth Particle Hydrodynamics (SPH) simulations (Monaghan 1992, Bate et al. 2003) using the DRAGON
SPH code from the Cardiff institute (see Goodwin et al. 2004). We simulated
the time evolution of an isothermal (T$_k=20$~K), Jeans-unstable elongated clump 
of mass 1000~M$_{\odot}$, comparable to the estimated total mass of NGC~2264-C 
($\sim 1600\, M_{\odot}$ -- PAB06). The model clump was initially ellipsoidal 
(finite boundary conditions) with an aspect ratio of 2.
The initial density profile was such as:
\begin{equation}
 n_{H_2} = \frac{n_c}{(1+(r/r_0)^2+(z/2 r_0)^2)}
\label{density}
\end{equation}
 corresponding to a flat inner ($r<r_0$) region and a $n_{H_2}\propto r^{-2}$ outer ($r>r_0$) region. The total mass of the flat inner region was $\sim 200$~M$_{\odot}$. Highly concentrated clouds (with, e.g., n$_{H_2} \propto r^{-2}$) are known to hardly fragment during collapse (cf. Myhill \& Kaula 1992; Whitworth et al. 1996), while uniform or moderately concentrated clouds 
(with n$_{H_2} \propto r^{-1}$ or flatter) typically fragment into as many 
Jeans-mass fragments as they contain initially (cf. Burkert et al. 1997). 
The expected number of fragments produced by the collapse 
of our model clump thus corresponds to the number of Jeans masses estimated 
in the flat inner core. At the mean density calculated within r$_0$ (i.e. n$_{H_2} \sim 1\times10^3$~cm$^{-3}$) and for T$_k = 20~K$, the Jeans mass 
is M$_J \simeq 20$~M$_{\odot}$ (see Bonnell et al. 1996
for a precise definition of M$_J$), which yields a Jeans mass 
number $N_J \sim 10$ for the flat inner core.
In these simulations we also included turbulent fluctuations. Since the exact nature and 
properties of interstellar turbulence are not fully understood yet, we 
considered two types of energy spectrum: 1) a spectrum scaling 
as Kolmogorov turbulence, i.e., $E(k)\propto k^{-5/3}$,  
and 2) a white spectrum, i.e, $E(k)\propto k^{0}$.
The phases of these turbulence fluctuations were chosen randomly.
Initially, three energy components controlled the evolution of the model 
filament: the thermal energy, $\mathcal T_{th}$, which remained constant 
in time throughout the simulations (i.e., isothermal assumption); 
the gravitational energy, $\mathcal W$, whose initial value depended on 
the clump density profile and thus on $n_c$ and $r_0$   
(larger values of $n_c$ and/or $r_0$ correspond to lower gravitational energy;
cf. Eq.(\ref{density})); and the turbulent energy, $\mathcal T_{turb}$.  
In all the simulations shown in this article, $n_c$ and $r_0$ 
have the same value, namely $n_c(H_2)=2000$~cm$^{-3}$ and $r_0=0.7$~pc, corresponding to an initial thermal to gravitational energy ratio $\alpha_{th}^0 \sim 8 \%$. 
The initial value of the ratio of turbulent to gravitational energy, $\alpha_{turb}^0$, 
was varied  from  0$\%$ to  50$\%$. We have also explored other initial conditions. When n$_c$ is too large ($n_c> 5000$~cm$^{-3}$), we find that too many fragments form, when n$_c$ is too small ($n_c< 500$~cm$^{-3}$), we find that only one central fragment will generally form. We have also varied the initial aspect ratio and conclude that if it is too close to one, then the cloud is not sufficiently filamentary whereas it is too much filamentary if it is initially too large. Finally, we have also explored the possibility that NGC~2264-C could be the result of a collision between two preexisting clouds rather than a collapsing elongated clump. However, it was not possible to reproduce the various features of this cloud within the scope of this scenario. The most important disagreements are i) a collision quickly tends to create a 
sheet rather than a filamentary object; ii) we find it very difficult to 
produce a well defined third object like C-MM3 by interaction of two colliding 
clouds; iii) a simple collision is unable to create a series of 
young condensations spread over the long axis of the clump at, e.g., 
positions comparable to C-MM1 and C-MM5. All our simulations were performed with a total of 5 million SPH particles.
 When the local density exceeded n$_{H_2} = 1.3 \times 10^8$ cm$^{-3}$, standard SPH particles were replaced by sink particles. The radius, $r_{sink} = 500~AU$, of the sink particles defines the highest resolution reached by our simulation. All particles falling within $r_{sink}$ of a sink particle and being bound to it were removed from the simulations, and their mass, linear and angular momentum were added to the corresponding sink particle values. Using sink particles allowed us to avoid artificial fragmentation
 (Truelove et al. 1997, Bate \& Burkert 1997). 
The relatively low density threshold at which sink particles were introduced 
implies that we could not model advanced phases of star/cluster formation but only the first stages of 
clump fragmentation.
Indeed, the limited numerical resolution of our simulations prevented us from describing
small spatial scale processes such as disk formation.

\subsection{General Pattern} 

Figure~\ref{sim} displays the density and velocity fields along the z-axis (i.e., long axis), as well as the column density maps in the (z,x) plane, 
at four time steps for a model filament with an initial level of turbulence, $\alpha_{turb}^0=5\%$. 
The reference time was chosen to be at one global free-fall time, t$_{ff}$, after the start of the simulation. 
Given the initial central density of the model clump, this corresponds to t$_{ff} = 9.5\times10^5$~yr. 
The four time steps shown in Fig.~\ref{sim} were taken at t$_{ff} - 4\times10^5$~yr,  t$_{ff}- 2\times10^5$~yr, 
t$_{ff}-1.0\times 10^{5}$~yr, and  t$_{ff}-0.4\times 10^{5}$~yr, respectively. 

We can describe the evolution of the model clump as follows (see also Bonnell et al. 1996; Inutsuka \& Miyama 1997).
Since the ellipsoidal clump initially contains several thermal Jeans masses (i.e. N$_J \sim 10$, cf. Sect.~3.1) and has a shorter dynamical timescale 
perpendicular to its major axis, it first collapses along its minor axis when seen projected onto the plane of the sky 
(see first and second panels of Fig.~\ref{sim}), 
amplifying the initial anisotropy and leading to the formation of a very elongated and filamentary structure (cf. Lin et al. 1965).
This fast contraction proceeds until thermal and turbulent pressure gradients 
can stop the collapse along the minor axis, ensuring an approximate hydrostatic equilibrium 
in the (xy) plane (cf Bonnell et al. 1996).

Since the dynamical timescale is longer along the major axis, the clump 
keeps collapsing along its long axis, i.e. the z-direction, after a transverse equilibrium has been 
established. 
The velocity field is initially nearly homologous 
(i.e., $V_z \propto z$ -- see first and second panels of Fig.~\ref{sim}) but becomes more and more 
complex as the filament fragments into several cores, each of them collapsing individually 
as can be seen on the third and fourth panels of Fig.~\ref{sim}. 
The individual collapse of the cores leads to the formation of local 
protostellar accretion shocks and associated protostars.
Moreover, due to the global collapse of the model clump toward its center, 
we can also see the formation of a central shock which separates the eastern ($z >0$) 
and western ($z <0$) sides of the clump. Altogether this dynamical 
evolution produces a complex density and velocity pattern.

When $\alpha_{turb}^0=50\%$, the velocity field (not displayed here for conciseness) is much less organized.
More shocks develop which lead to enhanced clump fragmentation through the process now widely 
referred to as ``turbulent fragmentation'' in the literature (e.g. Padoan \& Nordlund 2002, Klessen et al. 2005).
The number of shocks is larger and the model clump becomes more substructured as the initial level of turbulence 
increases (see also Jappsen \& Klessen 2004). 
At the other extreme, the case with no initial turbulence at all, $\alpha_{turb}^0=0\%$, does not yield any fragmentation 
due to the lack of initial fragmentation seeds, which is not realistic.

Therefore, based on the results of our simulations, several general conclusions can be drawn.
In particular, the initial level of turbulence appears to play a key role for the global aspect of the
model clump. The higher the turbulence, the more dispersed and less filamentary the model clump is. 
For low levels of initial turbulence, i.e., low values of $\alpha_{turb}^0$, the structure and kinematics 
of the clump are dominated by gravity, while for high levels of turbulence the clump is primarily structured 
by turbulence.

\section{Detailed comparison between the observations and the SPH simulations}

We performed a wide set of SPH simulations with different initial parameters, e.g., different values of the initial level
of turbulence, $\alpha_{turb}^0$, and of the thermal to gravitational energy ratio, $\alpha_{th}^0$. 
We did not find it necessary to use different initial turbulent velocity fields since 
the time evolution of the model clump depends only weakly on this. 
When calculating synthetic observations, we varied the inclination angle of the long axis of the model clump 
with respect to the line of sight. An inclination angle of 45 degrees was adopted to produce the synthetic images 
and diagrams shown in Fig.~\ref{comp_coldens} and Fig.~\ref{pv_diagrams}.
For each set of simulations, we seeked a particular time step at which the synthetic data, 
when convolved to the resolution of the observations, best matched the existing 30m and PdBI constraints.  
In the next subsection, we present our ``best-fit'' simulations and 
discuss the consequences of changing the best-fit parameters.

\subsection{Overall morphology: A fragmented filament}

\begin{figure*}[!ht!]

\hspace{0.0cm}
\includegraphics[height=18cm,angle=270]{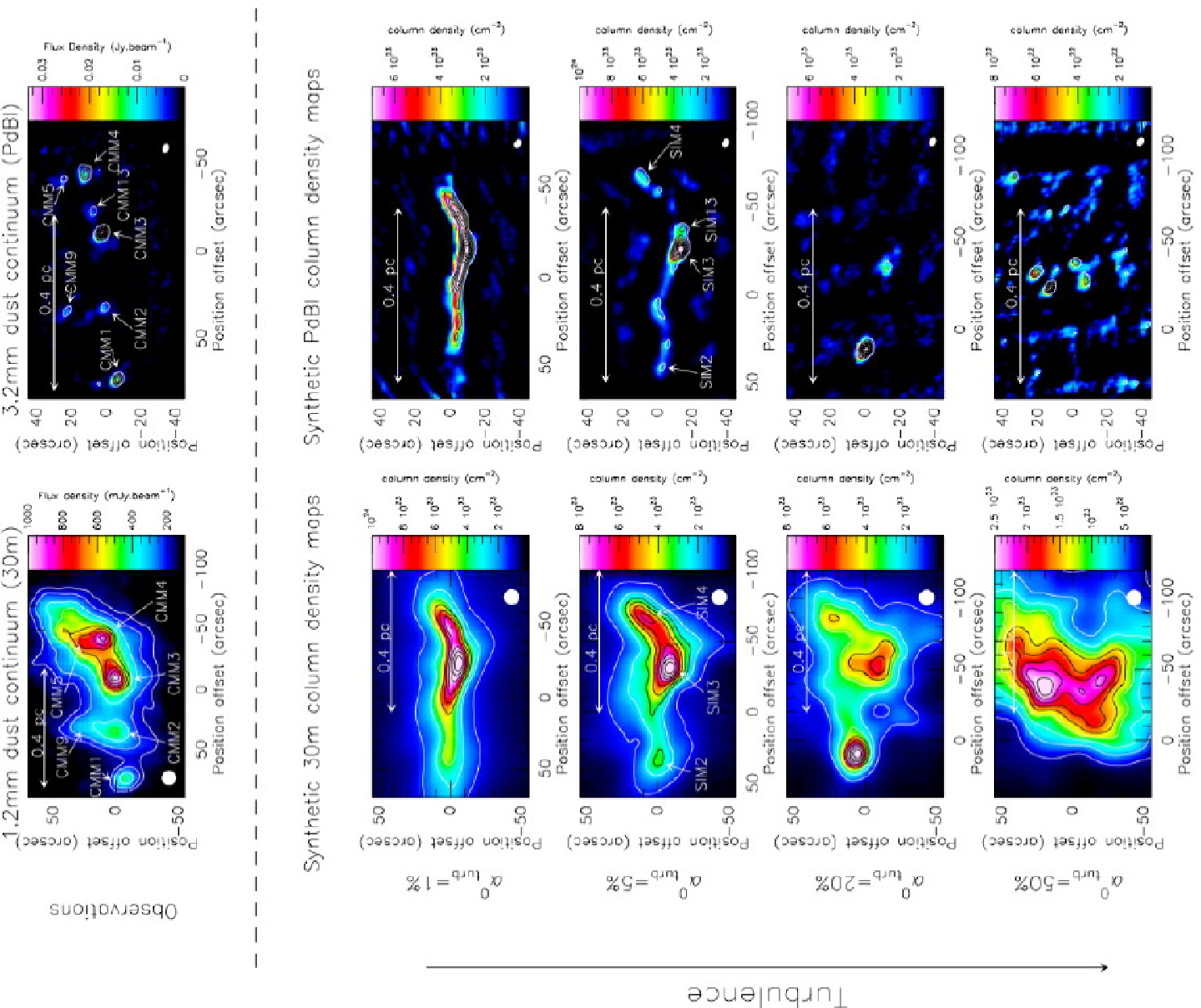}
\vspace{-0cm}
\caption{Observed column density distribution (first row) compared to synthetic column density maps (second to fifth row) 
convolved to the 30m angular resolution (first column) and to the PdBI angular resolution (second column) for four different initial levels of turbulence: $\alpha_{turb}^0=1\%$ (second row), $\alpha_{turb}^0=5\%$ (third row), $\alpha_{turb}^0=20\%$ (fourth row), $\alpha_{turb}^0=50\%$ (fifth row). 
The best-fit simulation corresponds to the third row, i.e., $\alpha_{turb}^0=5\%$  (for which the displayed time step is the ``best-fit'' time step). 
Note that the synthetic PdBI maps include the effect of interferometric filtering (see text). 
In each map, the contour levels go from 10 to 90$\%$ by step of 10$\%$ of the peak emission. 
The observed 30m column density distribution corresponds to the 1.2mm dust continuum map of PAB06.
\label{comp_coldens}}
\end{figure*}

The first important feature which must be reproduced by the simulations is the elongated shape of NGC~2264-C 
and the presence of several protostellar sources, lining up along the long (East-West) axis of the clump.

Figure \ref{comp_coldens} compares the observed column density maps (first row) with synthetic maps obtained from
simulations with four different initial levels of turbulence, i.e., $\alpha_{turb}^0=1\%$ (second row), $\alpha_{turb}^0=5\%$ (third row), $\alpha_{turb}^0=20\%$ 
(fourth row), and $\alpha_{turb}^0=50\%$ (fifth row), all assumed to be "observed" with a viewing angle of 45 degrees. 
The first column of Fig.~\ref{comp_coldens} displays the synthetic data convolved to the 30m resolution
while the second column displays the data convolved with the PdB interferometer beam. 
When convolving the simulated data to the PdBI resolution, we included the effect of 
interferometric filtering so as to allow more direct comparison with the observations. For this purpose, we used the UV\_MODEL task of the GILDAS package. This task generated a set of visibilities in the UV plane by calculating the values of the Fourier transform of the simulated input image at the observed UV baselines. 
The rest of the points in the UV plane was set to zero. 
This method had the consequence of filtering out all extended structures present in the numerical simulations. 

The four simulations shown in Fig.~\ref{comp_coldens} are compared when the synthetic column density maps convolved to the 30m resolution match the observed map best, except for the case with  $\alpha_{turb}^{0} = 50 \%$ (cf. fifth row) which does not exhibit a filamentary shape at any time step in the simulation. 
The corresponding time steps all lie in the range between t$_{ff} - 5\times10^5$~yr and t$_{ff} - 1.5\times10^5$~yr. 
The total mass accreted onto sink particles at these time steps ranges from 0$\%$ to 0.2$\%$ of the initial clump mass.
As already mentioned, we are thus looking at the very first stages of the formation of a protocluster, i.e., 
when the first pre-/proto-stellar cores with typical mean volume densities $\sim 10^5$~cm$^{-3}$ appear.
It can also be seen in Fig.\ref{comp_coldens} that the filamentary, elongated appearance of the NGC~2264-C clump 
cannot be reproduced when the initial level of turbulence is too high in the model.
Based on this argument, we conclude that $\alpha_{turb}^0$ has a maximum value of $20 \%$, 
although the $5\%$ model already provides a better match to the observations than the $20\%$ model. 
Figure~\ref{comp_coldens} also shows that the large-scale 
morphology of the clump observed at the resolution of the 30m telescope provides the 
strongest discriminator between different values of $\alpha_{turb}^0$.

Our ``best-fit'' simulation is shown in the third row of Fig.~\ref{comp_coldens}. 
It corresponds to a flat energy spectrum, $E(k)\propto k^{0}$, and 
an initial value of $\alpha_{turb}^0=5\%$, which is much lower 
compared to other numerical SPH studies of cloud 
fragmentation (e.g. Bate et al. 2003).  
Note that the case of Kolmogorov-like turbulence leads to results which are broadly 
similar to that shown in Fig.~\ref{comp_coldens}, except that the shape of the 
filament is more irregular and too distorted to match the observations well.
We therefore restrict our attention to the $E(k)\propto k^{0}$ case. Although this energy spectrum differs 
from the classical Kolmogorov one, we argue in Sect.~5 that it is not unrealistic on the parsec scale of 
the NGC~2264-C clump.

Comparison between the observed dust continuum maps of NGC~2264-C and the synthetic column density maps 
of the ``best-fit'' simulation (see Fig.~\ref{comp_coldens}) shows that the number of fragments and 
their alignment are well reproduced. By analogy with the observations, 
we have labelled 
the three main fragments of the synthetic 30m column density map SIM2, SIM3 and SIM4. 
The corresponding synthetic PdBI map shows a strong central source, SIM3, surrounded by weaker sources, as observed. 
Moreover an additional component, labelled SIM13, becomes visible next to SIM3 in the simulations when ``observed'' 
at the PdBI angular resolution, which is strongly reminiscent of the (C-MM3, C-MM13) system in the real interferometric map.

As described in \S~3.2, the collapse of the filament proceeds in two main phases: first, a global contraction velocity field 
is established along the long axis; second, a strong shock is generated at the center by the two interacting sides of the model clump. 
At the same time, the clump fragments
to form protostars. Thus, there are at least two relevant dynamical timescales in the problem: a global dynamical timescale corresponding to 
the global evolution of the elongated clump, and a local dynamical timescale corresponding to the dynamical
evolution of individual fragments. In other words, there is  
competition between local collapse (i.e. fragmentation) and global collapse. In our simulations, this
is controlled by the ratio of thermal to gravitational energy, $\alpha_{th}^0$, and thus by $n_c$ and r$_0$ 
(whose values are 2000 cm$^{-3}$ and 0.7~pc, respectively) since 
the kinetic temperature and the mass of the model clump are fixed (cf \S~3.1). 
If $n_c$ or r$_0$ are too small (i.e. the density structure approaches $n_{H_2} \propto r^{-2}$; cf Eq.(\ref{density})), 
the individual fragments do not have enough time to collapse on their own before entering the 
central shock. Therefore only one, massive central core forms. 
Conversely, if $n_c$ and r$_0$ are too large, many protostars
(and eventually stars) form before any significant large-scale velocity field
is established along the clump long axis. 
Furthermore, in the latter case, the number of fragments produced in the simulations 
becomes larger than the observed number of fragments.

Note that  the collapse simulations shown in this paper were performed with model clumps of total mass $M_{tot} = 1000\, M_{\odot}$, 
while the total mass of NGC~2264-C is estimated to be somewhat larger,  $\sim 1600$~M$_{\odot}$ (cf. PAB06). 
With more massive model clumps, we did not manage to reproduce the overall morphology of NGC~2264-C, 
in the sense that too much fragmentation occurred. 
This suggests that our models lack some source of support against gravity compared to the actual NGC~2264-C clump.
This will be discuss further in \S~5.

\subsection{The sharp central velocity discontinuity}

\begin{figure*}[!th!]
\hspace{1cm}
\includegraphics[height=15cm,angle=270]{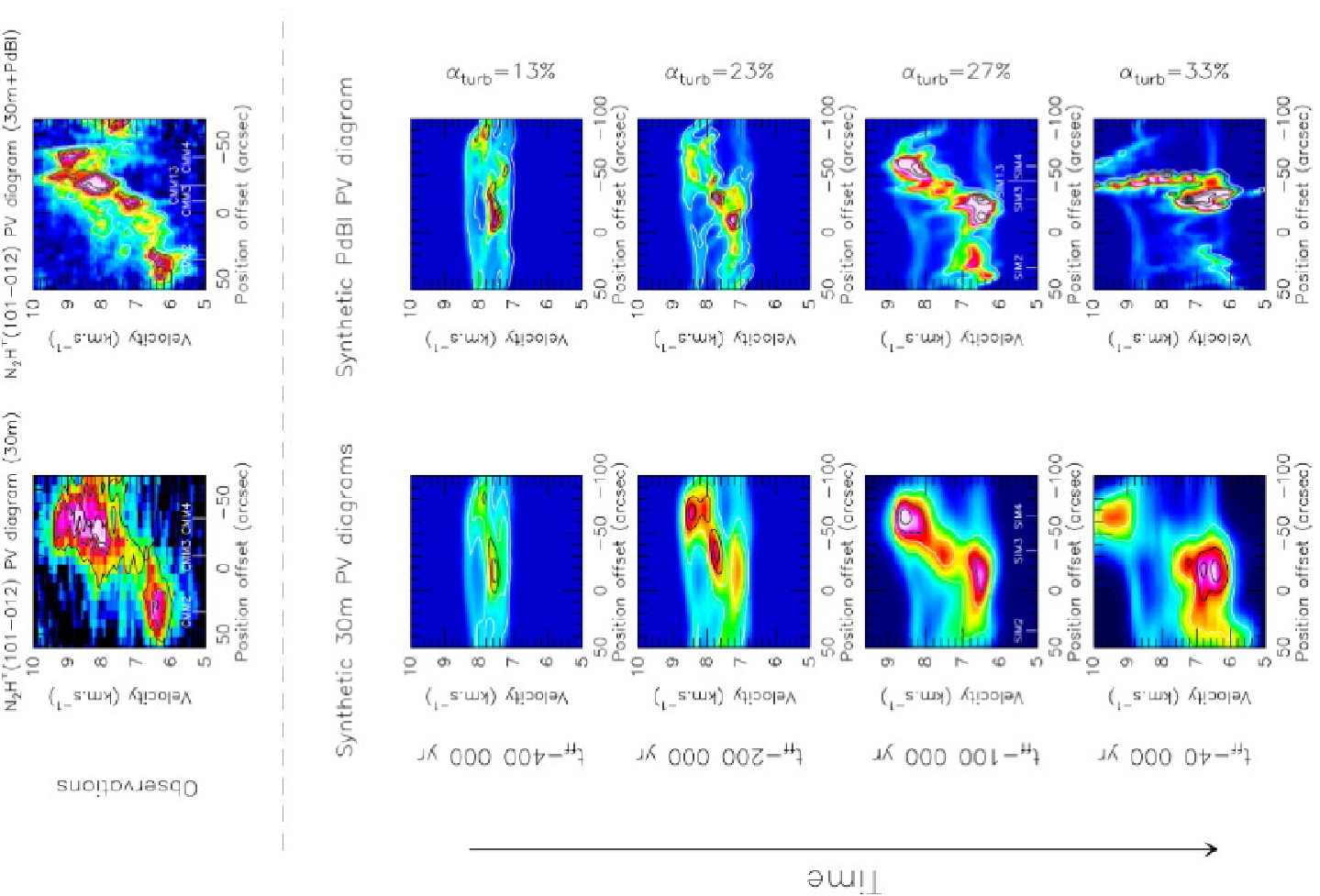}
\caption{Observed position-velocity diagrams (first row) compared to synthetic position-velocity diagrams (second to fifth row) convolved to the 30m angular resolution (first column) and to the PdBI angular resolution (second column) at four different time steps of our best-fit SPH simulation ($\alpha_{turb}^0 = 5\%$): t$_{ff}-400 000$~yr (second row); t$_{ff}-200 000$~yr (third row); t$_{ff}-100 000$~yr (fourth row); t$_{ff}-40 000$~yr (fifth row) (with t$_{ff}=9.5\times 10^5$~yr). The best-fit time step corresponds to fourth row. 
The value of the turbulent to gravitational energy ratio $\alpha_{turb}$ at each time step is given on the right hand side. 
Note that $\alpha_{turb}$ increases as time proceeds in the simulation.
\label{pv_diagrams}}
\end{figure*}

One of the most interesting features of NGC~2264-C is the central velocity 
discontinuity observed by PAB06 in optically thin tracers toward C-MM3 (see Fig.~\ref{pv_diagrams}). 
This velocity discontinuity is believed to trace the axial (i.e., 1D) collapse of NGC~2264-C along
its long axis, as well as a possible dynamical interaction between protostellar sources at the center of the clump. 
Our new PdBI observations, which confirm the presence of a strong velocity gradient along the long axis 
of the clump (i.e. $\sim 8.4$~km.s$^{-1}$.pc$^{-1}$ -- see Fig.~\ref{pv_n2h+}), set additional 
constraints on the velocity field in the central part of NGC~2264-C.

When the initial level of turbulence was lower than $20\%$, 
our SPH simulations convolved to the 30m resolution led to a central discontinuity resembling that observed. 
Furthermore, the shape of the PV diagram observed at the PdBI resolution along the long axis of the clump 
appears to be a key tracer of the time evolution, as can be seen in Fig.~\ref{pv_diagrams}. 
In addition to observed PV diagrams (first row), Fig.~\ref{pv_diagrams}
shows the synthetic PV diagrams of our ``best-fit'' simulation ($\alpha_{turb}^0 = 5\%$) convolved to the 30m resolution (first column) and to the PdBI resolution 
(second column) at the same four time steps as in Fig.~\ref{sim} (rows two to five). 
Note that simulations adopting a Kolmogorov-like turbulent energy spectrum and low values of $\alpha_{turb}^0$ ($\le 20\%$) 
led to velocity discontinuities which are similar to the discontinuity shown here for the ``best-fit'' simulation.

The reference time in Fig.~\ref{pv_diagrams} is the same as in Fig.~\ref{sim}, namely one global free-fall time  
(t$_{ff} = 9.5\times10^5$~yr) after the start of simulation. 
At the first time step shown, t$_{ff}-4\times 10^5$~yr (second row), no clear kinematical signature is apparent in the synthetic PV diagrams, either at the 30m or 
at the PdBI resolution. At  t$_{ff}-2\times 10^5$~yr (third row), the synthetic PV diagrams start to exhibit a velocity structure reminiscent of 
the observed velocity gradient and central discontinuity, but the amplitude of the velocity structure is not large enough to match the observations.
At the best-fit time step, i.e., t$_{ff}-1\times10^5$~yr,  
the agreement between the simulated PV diagrams (fourth row) and the observed PV diagrams (first row) is quite remarkable. 
The central amplitude (i.e. $\sim2$~km.s$^{-1}$), shape, and position of the velocity discontinuity 
are well reproduced. Moreover, the synthetic 
PdBI PV diagram shows a $\sim 1$~km.s$^{-1}$ velocity gap between the two
central fragments, SIM3 and SIM13, as observed between C-MM3 and C-MM13. 

The fifth row of Fig.~\ref{pv_diagrams} shows a later time, i.e., t$_{ff}-0.4\times 10^5$~yr, when the central shock is well developed. 
While the synthetic 30m PV diagram remains satisfactory, the synthetic PdBI PV diagram differs markedly from the observations. 
Note, however, that this late phase of evolution may not be correctly described by our numerical model. 
Indeed, our simulations do not include feedback from protostars 
which clearly influences the late dynamical evolution of cluster-forming clouds (cf. Li \& Nakamura 2006).
Thus, it is not clear if, in reality, the central shock would have 
time to develop as much as in the simulations to produce a PV diagram such as the one shown at the bottom right 
of Fig.~\ref{pv_diagrams}.

The time evolution of the $\alpha_{turb}$ ratio in the ``best-fit'' simulation is also given on the right-hand side of Fig.~\ref{pv_diagrams}.
While the initial level of turbulence was only $\alpha_{turb}^0=5\%$ in this simulation, it can be seen 
the ratio of nonthermal kinetic energy to gravitational energy quickly increases up to $\alpha_{turb}=27\%$ at the ``best-fit'' 
time step and $\alpha_{turb}=33\%$ at the last time step shown. 
This demonstrates that the bulk of the ``turbulent'' energy in our simulations does not come from the large scale turbulent velocity field, 
but rather from the conversion of gravitational energy into kinetic energy through the global collapse of the clump. 
The increase of $\alpha_{turb}$ with time contributes to produce synthetic linewidths in reasonable agreement with 
observed linewidths (see \S~4.3 and Fig.~\ref{comp_spec} below), despite the low level of kinetic energy at the beginning of the simulation. 
We also note that the value of $\alpha_{turb}$ achieved at the ``best-fit'' time step (27\%) is within a factor of 2 of 
the kinetic to gravitational energy ratio expected in virial equilibrium (50\%), despite the fact that the model clump is globally 
collapsing and far from equilibrium at this stage. Clearly, the broad linewidths observed in NGC~2264-C result at least partly 
from systematic inward motions as opposed to random turbulence.
In our  ``best-fit'' model, most of the motions are gravitationally focussed and do not exert any support against gravitational collapse. 

\subsection{Large-scale kinematical pattern in optically-thin line tracers}
\begin{figure*}[t]

\vspace{-0.0cm}
\hspace{2cm}
\includegraphics[height=11cm,angle=270]{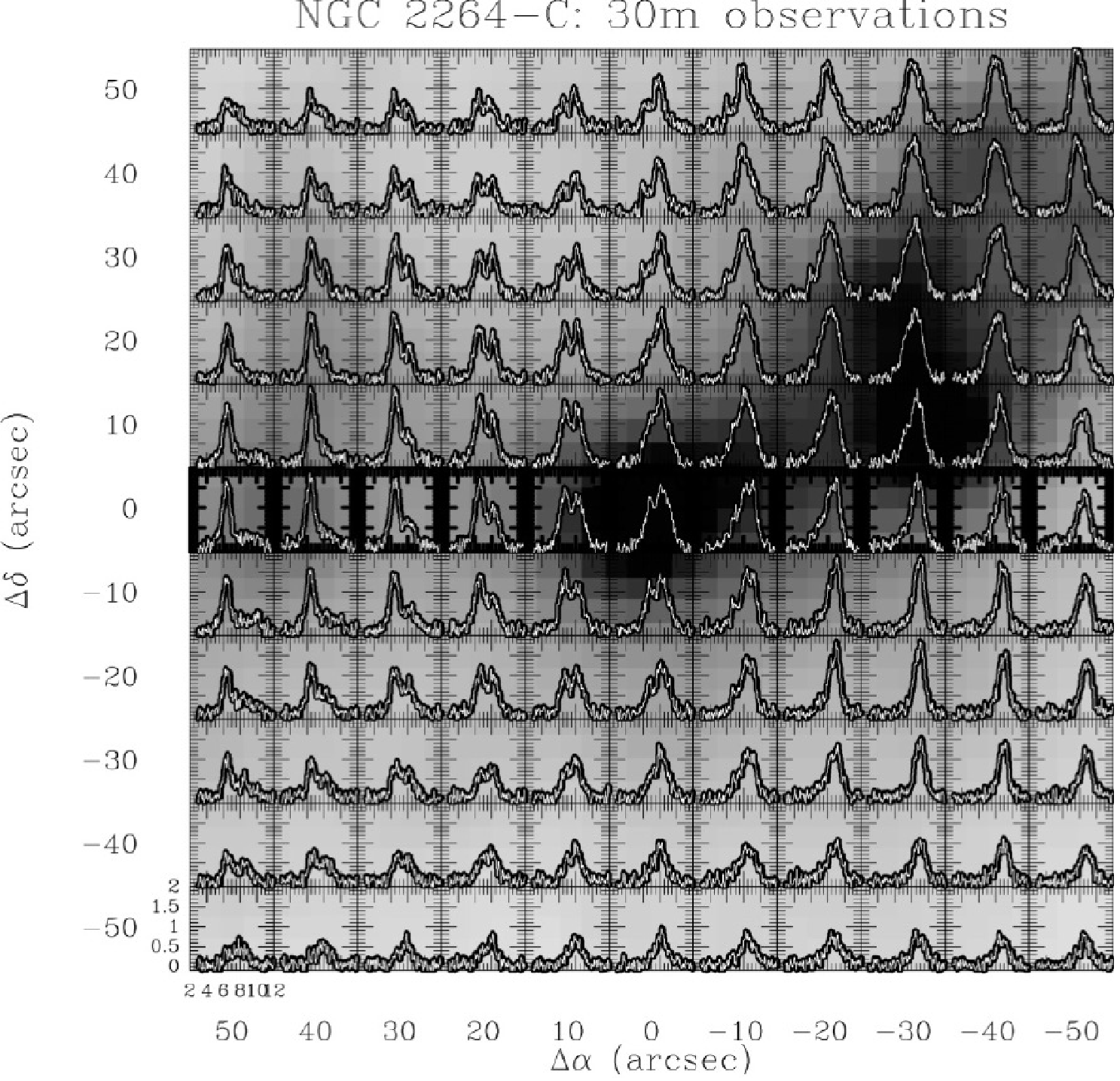}

\vspace{0.5cm}
\hspace{2cm}
\includegraphics[height=11cm,angle=270]{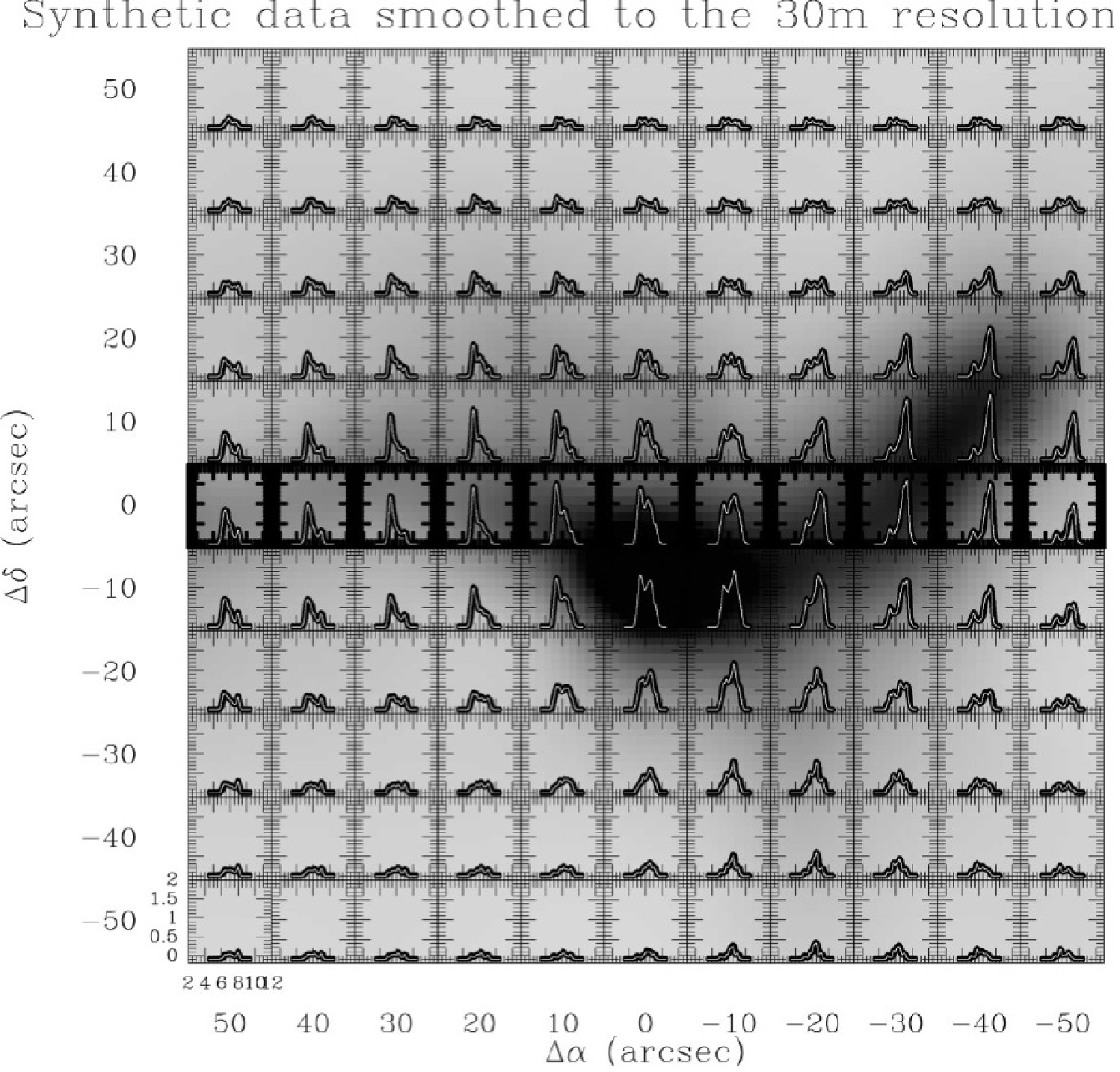}
\vspace{-0cm}
\caption{Comparison between the H$^{13}$CO$^{+}$(1-0) spectra observed at the 30m telescope (upper panel) and 
the synthetic optically thin spectra obtained in our best-fit simulation (lower panel).  
In the upper panel, the (0,0) corresponds to the position of C-MM3, while in the lower panel the (0,-10) position corresponds to the position of SIM3.
Overlaid in grey scale are the 1.2~mm dust continuum image of PAB06 (top) and the synthetic column density 
map of the best-fit simulation (bottom).
The row of spectra observed (top) and simulated (bottom) along the main axis of the clump are marked 
in boldface. 
\label{comp_spec}}
\end{figure*}

The H$^{13}$CO$^+$(1-0) spectra observed toward NGC~2264-C with the 30m telescope
show a remarkable East-West axial symmetry over the whole clump 
on either side of the central source C-MM3 (see Fig.~\ref{comp_spec}). 
The low optical depth of the H$^{13}$CO$^+$(1-0) line 
($\tau \sim 0.3$ for the peak velocity channel of the central spectrum) 
inferred from the Monte-Carlo radiative transfer calculations of PAB06 implies that the observed double-peaked line profiles
(cf. Fig.~\ref{comp_spec}) result from the presence of two velocity components 
along the line of sight rather than from self-absorption. 
Although our PdBI observations show that the central velocity discontinuity seen with the 30m telescope 
originates from two unresolved protostellar sources, the large spatial extent of the region over which 
double-peaked H$^{13}$CO$^+$(1-0) spectra are observed in the 30m map suggests a global kinematical origin 
for the double-peak profiles.

While the N$_2$H$^+$(101-012) PV diagram observed with PdBI (Fig.~\ref{pv_n2h+}) may be
suggestive of rotation about an axis perpendicular to the long axis of the clump, 
PAB06 showed that rotation could not account for the shape of the observed 30m PV diagram 
based on a detailed comparison with radiative transfer models (see Fig.~12 of PAB06).
By contrast, we now proceed to show that our scenario of large-scale, axial collapse does provides a good match 
to the symmetric pattern of double-peaked line profiles observed in low optical depth tracers.

To this aim, synthetic spectra were constructed from our SPH simulations assuming strictly optically thin
line tracers: each SPH particle was given the same weight and the contributions of 
all particles falling within a given velocity channel were integrated. 
The synthetic data cube was then convolved to the 30m angular resolution and 
normalized in such a way that the peak intensity of the synthetic central spectrum 
matched the peak intensity of the observed central H$^{13}$CO$^{+}$(1-0) spectrum. 
Figure~\ref{comp_spec} compares the H$^{13}$CO$^{+}$(1-0) spectra observed in the central part of NGC~2264-C 
(top) with the resulting synthetic spectra for the best-fit simulation at the best-fit time step (bottom).
(In this comparison, we use the observed H$^{13}$CO$^{+}$(1-0) spectra rather than the observed N$_2$H$^{+}$(1-0) spectra 
since the former have a better signal-to-noise ratio.) 
It can be seen that the overall agreement is very good. Since the synthetic line emission is optically thin,  
the double-peaked spectra exhibited by the model are clearly not due to radiative transfer effects 
but result from the presence of two velocity components along the line of sight, 
corresponding to the two ends of the elongated clump moving toward each other.
Focusing on the central row of spectra (marked in boldface), 
it can be seen that the blue-shifted component of the double-peaked spectra dominates on the eastern side. 
Moving west, the red-shifted component becomes progressively stronger. It is nearly as intense as the blue-shifted component 
at the central position and eventually dominates on the western side of the filament. 
This remarkable reversal of blue/red spectral asymmetry as one moves from the eastern to the western 
side of the central C-MM3 position can be seen in both the observations and simulations. 

The synthetic spectra obtained from our SPH simulations are mass weighted and are thus more representative 
of the global kinematics of the clump than of the kinematics of compact individual fragments. 
We conclude that the remarkable pattern seen in the central row of spectra in Fig.~\ref{comp_spec} 
characterizes the collapse of the elongated clump along its long axis. 
We note, however, that the synthetic spectra are somewhat narrower than are the observed H$^{13}$CO$^{+}$(1-0) spectra. 
Part of the observed linewidths may result from outflowing gas generated by the protostars, an effect which we did not treat in our 
simulations. It may also be partly due to another source of support against gravity, 
not included in our simulations (see \S ~5).

In the context of our interpretation of the double-peaked spectral pattern observed in H$^{13}$CO$^{+}$(1-0),  
the extent of the region over which the red-shifted peak is observed 
on the eastern side (and the blue-shifted peak observed on the western side) 
sets constraints on the diameter of the NGC~2264-C cylinder-like clump (see Fig.~11 of PAB06).
Double-peaked H$^{13}$CO$^{+}$(1-0) spectra are observed up to 30\arcsec 
~on either side of the central object C-MM3. Given the distance of 800~pc and assuming 
a viewing angle of 45 degrees between the line of sight and the long axis of the clump 
(as adopted in the radiative transfer model presented by PAB06), we estimate the diameter 
of the cylinder to be $\sim 0.65$~pc. 
This is in good agreement with the apparent width of the NGC~2264-C clump as measured 
in the plane of sky on our dust continuum and molecular line maps.

\section{Concluding remarks}

The good quantitative agreement obtained between our ``best-fit'' SPH simulations and 
our (30m and PdBI) millimeter observations confirms the physical plausibility of the 
scenario of large-scale axial collapse and fragmentation proposed by PAB06 for the NGC~2264-C clump. 
Observationally, such an axial collapse is traced by a central velocity 
discontinuity associated with double-peaked profiles in optically thin line tracers. 
The present study supports our earlier suggestion that an ultra-dense protostellar core of mass 
up to $\sim 90\, M_\odot $ is in the process of forming at the center of NGC~2264-C through the 
dynamical merging of lower-mass Class~0 cores (cf. PAB06). 
Our interferometric PdB detection of a new object, C-MM13, located only $\sim 10000$~AU away
(in projection) from the central source, C-MM3, but with a line-of-sight velocity differing by 
$\sim 1.1$~km.s$^{-1}$ from that of C-MM3, provides an additional observational manifestation 
of the merging process. Given the relatively large mass of C-MM13 ($\sim 8\, M_\odot$), 
such a large velocity difference would be difficult to explain by dynamical fragmentation during 
the collapse of an individual protostellar core, even if low-mass objects can easily be ejected from
dynamically unstable protostellar systems (e.g. Bate et al. 2003, Goodwin et al. 2004).
In our proposed scenario for NGC~2264-C, the local collapse of individual protostellar cores is strongly influenced by the high 
dynamical pressure resulting from the global collapse of the clump, and proceeds in a manner that is 
qualitatively similar to the triggered protostellar collapse models discussed by Hennebelle et al. (2003, 2004).

Our detailed comparison between observations and simulations has also allowed us to
set constraints on the evolutionary state of the NGC~2264-C protocluster. 
It seems that the characteristic shape of the observed position-velocity diagrams survives 
for a relatively short period of time, i.e. $\le 1\times10^{5}$~yr, 
and occurs only very soon after the formation of the protocluser
while less than $\sim 1\%$ of the gas has been accreted onto sink particles. 

The low level of  initial turbulent energy required to match the observations
implies that NGC~2264-C is structured more by self-gravity than by turbulence.
The main effect of turbulence is to create seeds for further gravitational fragmentation. 
Turbulent fragmentation does not appear to play a significant role in this clump.
In our ``best-fit'' simulation,  the initial turbulent to gravitational energy ratio is 
$\alpha_{turb}^0=5\%$, comparable to the ratio of thermal to gravitational energy $\alpha_{th}^0$. 
The level of ``turbulence'' increases as the simulation proceeds and gravitational energy 
is converted into kinetic energy. At the ``best-fit''  time step, the ratio of nonthermal kinetic energy 
to gravitational energy approaches $\sim 30\% $. Most of the corresponding ``turbulence'' is 
gravitationally generated as in the recent cloud collapse simulations of Burkert \& Hartmann (2006).
In other words, the cloud motions in our best-fit model are primarily due 
to collapse and gravitationally organized motions as opposed to purely random turbulence. 
Although we have not identified a specific trigger, we believe that the ``cold'' or ``subvirial'' 
initial conditions (cf. Adams et al. 2006) required by our model reflect the fact that the NGC~2264-C clump 
was suddenly compressed and/or assembled as a result of a strong external perturbation.

The fact that simulations starting from initial turbulent velocity fields 
with a Kolmogorov-like energy spectrum lead to model clumps that are much less organized 
than the observations should not be overinterpreted. 
Indeed, since the phases are chosen randomly and since in Kolmogorov-like turbulence 
most of the energy is on large scales, it is 
not surprising that the shape of the filament is strongly distorted in this case. 
In a real situation, the large-scale turbulent fluctuations should be much more 
coherent since they may be responsible, at least in part, for the 
formation of the filament in the first place (cf. Hartmann et al. 2001).

Another point worth noting is that the total mass of gas with density above 
$10^{4}$~cm$^{-3}$ is $\sim 10$ times lower in our best-fit simulation 
than in the actual NGC~2264-C clump.
Using higher densities by a factor of 10 in the numerical simulations 
would inevitably lead to fragmentation into a larger number of cores since the corresponding Jeans
mass would be smaller by a factor $\sim 3$ compared to the Jeans mass in the present simulations. 
It seems therefore that some additional support against gravity, not included in the simulations 
presented here, plays a role in NGC~2264-C. This extra support could arise from 
protostellar feedback and/or magnetic fields.

Finally, we speculate that the evolution inferred and simulated here for NGC~2264-C is not 
exceptional but representative of many massive cluster-forming clumps in the Galaxy. 
In particular, we note that evidence of large-scale, supersonic inward motions has been 
recently found in several deeply embedded regions of high-mass star formation 
(Motte et al. 2005 -- see also Wu \& Evans 2003 and Fuller et al. 2005).
NGC~2264-C may just be caught at a particularly early stage of protocluster evolution 
and observed in a favorable configuration, leading to a remarkably simple kinematical 
pattern. Similar detailed modelling studies of other cluster-forming clumps will be needed to confirm 
this hypothesis.

\begin{acknowledgements} 
We are grateful to the IRAM astronomers in Grenoble for their help 
with the Plateau de Bure interferometric observations. 
IRAM is supported by INSU/CNRS (France), MPG (Germany), and IGN (Spain). 

\end{acknowledgements}

\end{document}